\documentclass{ifacconf}

\usepackage{graphicx}      % include this line if your document contains figures
\usepackage{natbib}        % required for bibliography
\usepackage{tikz}
\usetikzlibrary{calc,patterns,
                 decorations.pathmorphing,
                 decorations.markings}
\usepackage{amsmath}
\usepackage{amssymb}
\usepackage{graphicx}
\usepackage{mathtools}
\begin{document}
\begin{frontmatter}

\title{Damage Modeling for the Tree-Like Network with Fractional-Order Calculus\thanksref{footnoteinfo}} 

\thanks[footnoteinfo]{The partial support of the US NSF Award 1826079 is gratefully acknowledged.}

\author[First]{Xiangyu Ni} 
\author[Second]{Bill Goodwine} 

\address[First]{Department of Aerospace and Mechanical Engineering, University of Notre Dame, IN 46556 USA (e-mail: xni@nd.edu).}
\address[Second]{Department of Aerospace and Mechanical Engineering, University of Notre Dame, IN 46556 USA (e-mail: bill@controls.ame.nd.edu)}

\begin{abstract}                % Abstract of not more than 250 words.
In this paper, we propose that a tree-like network with damage can be modeled as the product of a fractional-order nominal plant and a fractional-order multiplicative disturbance, which is well structured and completely characterized by the damage amount at each damaged component. Such way of modeling brings us insights about that damaged network's behavior, helps us design robust controllers under uncertain damages and identify the damage. Although the main result in this paper is specialized to one model, we believe that this way of constructing a well-structured disturbance model can be applied to a class of damaged networks.
\end{abstract}


\end{frontmatter}
%===============================================================================

\section{INTRODUCTION}

Controlling and health monitoring large networks are research topics having a long history. For example, see the survey paper \cite{murray2007recent} and the book \cite{ren2008distributed} for multi-vehicle cooperative control, and the papers \cite{CAO2010233, 4140748} for formation control. For health monitoring, see the papers \cite{doi:10.1002/stc.290} and \cite{worden2008review}. Our approach in this paper extends this body of knowledge by leveraging fractional-order calculus which allows us to exactly model the damage inside a large network. Then, using that knowledge from modeling, our ultimate goal is to design robust controllers and to identify damages for such large networks.

This paper studies a specific network shown in Fig.~\ref{fig:tree}, motivated by a viscoelastic model from \cite{heymans1994fractal}. If we limit ourselves to integer-order calculus, that system can only be modeled by an infinite continued fraction. Existing literature, for example \cite{goodwine2014modeling} shows that, if fractional-order calculus is allowed, then the undamaged version of that system is exactly half order which has a very concise representation. However, to our knowledge, no existing literature shows how to model that network when it is damaged. Therefore, in the paper, we show that even when a damage exists in such network, its analytical transfer function can still be written in a well-structured and concise way.

In addition to real network systems, we believe that this research could have more impact on other systems governed by fractional-order differential equations. For example, see \cite{heymans1994fractal} which proposes that the network shown in Fig.~\ref{fig:tree} can be a rheological model of viscoelastic behavior which obviously does not contain any real networks. However, the reason why network models still appear in that research area is exactly because its response is of fractional-order. This is the reason why we do not believe an infinitely large network such as the one shown in Fig.~\ref{fig:tree} is for purely academic purpose. That is to say, although an infinitely large network does not actually exist, the fractional-order differential equation describing its dynamics can still make it useful for some real systems.

Research in fractional-order calculus dates back almost to the birth of calculus. In recent decades, it has gained an increased focus with a larger number of applications to complicated systems due to its intrinsic properties. First of all, fractional-order derivatives are non-local, so they are used to model epidemics: \cite{ahmed2007fractional}. Moreover, the time-domain response for a linear fractional-order system can follow a power-law decay rate. That property leads to a modeling example of the firing rate for premotor neurons in our visual system while our eyeball is scanning words as shown in Chapter 1 of \cite{magin2006fractional}.

The rest of this paper is organized as follow. In Section \ref{sec:background}, we give some background information about determining the dynamics of the network shown in Fig.~\ref{fig:tree}. In Section \ref{sec:mathAnalysis}, we give the main result of this paper, that is modeling the damaged version of that network in Fig.~\ref{fig:tree}, and prove that result. In Section \ref{sec:discussion}, we discuss some applications which leverage the advantages brought by our way of modeling. Finally, Section \ref{sec:conclusion} gives concluding remarks and our future research focus in regard to this topic.

\section{BACKGROUND}
\label{sec:background}

\subsection{Tree Model}

\begin{figure}
\centering
\begin{tikzpicture}
\tikzstyle{spring}=[thick,decorate,decoration={zigzag,pre length=0.3cm,post length=0.3cm,segment length=6}]
\tikzstyle{damper}=[thick,decoration={markings,mark connection node=dmp,mark=at position 0.5 with 
   {
    \node (dmp) [thick,inner sep=0pt,transform shape,rotate=-90,minimum
 width=15pt,minimum height=3pt,draw=none] {};
    \draw [thick] ($(dmp.north east)+(2pt,0)$) -- (dmp.south east) -- (dmp.south
 west) -- ($(dmp.north west)+(2pt,0)$);
    \draw [thick] ($(dmp.north)+(0,-5pt)$) -- ($(dmp.north)+(0,5pt)$);
   }
 },decorate]
 
\node at (0,0) (leftF) {$f$};
\filldraw[black] (0.7,0) circle (2pt) node [above] {$x_{1,1}$};
\draw[thick, -latex] (leftF.east) -- (0.7,0);
\draw[thick] (0.7,0) -- (1.1,0);
\draw[thick] (1.1,1) -- (1.1,-1);
\draw[spring] (1.1,1) -- (2.6,1) node [midway,above=1pt] {$k_{1,1}$};
\draw[damper] (1.1,-1) -- (2.6,-1) node [midway,above=6pt] {$b_{1,1}$};
\draw (2.6,1) circle (2pt) node [above] {$x_{2,1}$};
\draw (2.6,-1) circle (2pt) node [above] {$x_{2,2}$};
\draw[thick] (2.6,1) -- (3,1);
\draw[thick] (2.6,-1) -- (3,-1);
\draw[thick] (3,1.5) -- (3,0.5);
\draw[thick] (3,-0.5) -- (3,-1.5);
\draw[spring] (3,1.5) -- (4.5,1.5) node [midway,above=1pt] {$k_{2,1}$};
\draw[damper] (3,0.5) -- (4.5,0.5) node [midway,above=6pt] {$b_{2,1}$};
\draw[spring] (3,-0.5) -- (4.5,-0.5) node [midway,above=1pt] {$k_{2,2}$};
\draw[damper] (3,-1.5) -- (4.5,-1.5) node [midway,above=6pt] {$b_{2,2}$};
\draw (4.5,1.5) circle (2pt) node [above] {$x_{3,1}$};
\draw (4.5,0.5) circle (2pt) node [above] {$x_{3,2}$};
\draw (4.5,-0.5) circle (2pt) node [above] {$x_{3,3}$};
\draw (4.5,-1.5) circle (2pt) node [above] {$x_{3,4}$};
\draw[thick] (4.5,1.5) -- (5,1.5);
\draw[thick] (4.5,0.5) -- (5,0.5);
\draw[thick] (4.5,-0.5) -- (5,-0.5);
\draw[thick] (4.5,-1.5) -- (5,-1.5);
\draw[thick] (5,2) -- (5,1);
\draw[spring] (5,2) -- (6.5,2) node [midway,above=1pt] {$k_{3,1}$};
\draw[damper] (5,1) -- (6.5,1) node [midway,above=6pt] {$b_{3,1}$};
\draw (6.5,2) circle (2pt) node [above] {$x_{4,1}$};
\draw (6.5,1) circle (2pt) node [above] {$x_{4,2}$};
\node at (7,2) {$\cdots$};
\node at (7,1) {$\cdots$};
\node at (5.5,0.5) {$\cdots$};
\node at (5.5,-0.5) {$\cdots$};
\node at (5.5,-1.5) {$\cdots$};
\draw[black, very thick] (7.3,2.5) rectangle (7.5,-2) node [midway,below=2.4cm] {$x_{last}$};
\filldraw[black] (7.4,2.4) circle (2pt);
\filldraw[black] (7.4,2.2) circle (2pt);
\filldraw[black] (7.4,2) circle (2pt);
\filldraw[black] (7.4,1.8) circle (2pt);
\filldraw[black] (7.4,1.6) circle (2pt);
\filldraw[black] (7.4,1.4) circle (2pt);
\node at (7.4,1.2) {$\vdots$};
\node at (7.4,0.7) {$\vdots$};
\node at (8.2,0) (rightF) {$f$};
\draw[thick, -latex] (rightF.west) -- (7.5,0);
\end{tikzpicture}
\caption{The tree model.}
\label{fig:tree}
\end{figure}
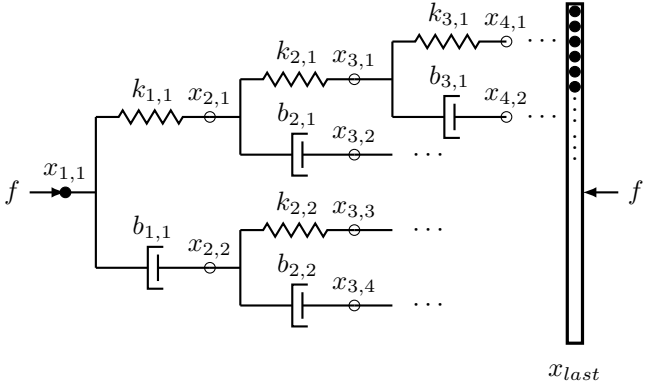

This paper mainly focuses on the tree model shown in Fig.~\ref{fig:tree}. This model has an infinite number of generations. Each generation doubles its number of nodes from the previous one. The upper node is connected to its previous generation through a spring, and the lower node is connected to its previous generation through a damper. The last generation's nodes are locked together. This type of system has been studied extensively in the fractional calculus literature. See, for example, \cite{heymans1994fractal,goodwine2014modeling,leyden2018system,mayes2012reduction}.

\begin{figure}
\centering
\begin{tikzpicture}
\tikzstyle{spring}=[thick,decorate,decoration={zigzag,pre length=0.3cm,post length=0.3cm,segment length=6}]
\tikzstyle{damper}=[thick,decoration={markings,mark connection node=dmp,mark=at position 0.5 with 
   {
    \node (dmp) [thick,inner sep=0pt,transform shape,rotate=-90,minimum
 width=15pt,minimum height=3pt,draw=none] {};
    \draw [thick] ($(dmp.north east)+(2pt,0)$) -- (dmp.south east) -- (dmp.south
 west) -- ($(dmp.north west)+(2pt,0)$);
    \draw [thick] ($(dmp.north)+(0,-5pt)$) -- ($(dmp.north)+(0,5pt)$);
   }
 },decorate]
 
\node at (0,0) (leftF) {$f$};
\filldraw[black] (0.7,0) circle (2pt) node [above] {$x_{1,1}$};
\draw[thick, -latex] (leftF.east) -- (0.7,0);
\draw[thick] (0.7,0) -- (1.1,0);
\draw[thick] (1.1,0.5) -- (1.1,-0.5);
\draw[spring] (1.1,0.5) -- (2.5,0.5) node [midway,above=1pt] {$\widetilde{k}$};
\draw[damper] (1.1,-0.5) -- (2.5,-0.5) node [midway,above=6pt] {$\widetilde{b}$};
\draw (2.5,0.5) circle (2pt) node [above] {$x_{2,1}$};
\draw (2.5,-0.5) circle (2pt) node [above] {$x_{2,2}$};
\node[draw,outer sep=0pt,thick] (M1) at (3.5,0.5) {$\widetilde{G}_U(s)$};
\node[draw,outer sep=0pt,thick] (M2) at (3.5,-0.5) {$\widetilde{G}_L(s)$};
\draw[thick] (2.5,0.5) -- (M1.west);
\draw[thick] (2.5,-0.5) -- (M2.west);
\draw[black, thick] (4.4,1.2) rectangle (4.5,-1.2) node [midway,below=1.2cm] {$x_{last}$};
\draw[thick] (M1.east) -- (4.4,0.5);
\draw[thick] (M2.east) -- (4.4,-0.5);
\node at (5.2,0) (rightF) {$f$};
\draw[thick, -latex] (rightF.west) -- (4.5,0);
\end{tikzpicture}
\caption{An illustration about the elements in the recurrence formula, Eq.~(\ref{eq:treeRecursive}).}
\label{fig:recursive}
\end{figure}
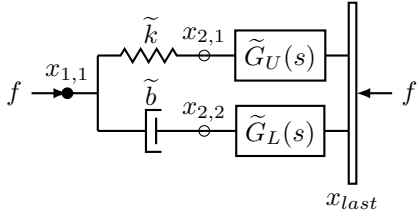

It can be shown that the transfer function $\widetilde{G}(s)$ from the input force, $f$, to the relative distance between $x_{1,1}$ and $x_{\text{last}}$ of such model satisfies the recurrence formula given by
\begin{equation}
    \widetilde{G}(s)=\cfrac{1}{\cfrac{1}{\cfrac{1}{\widetilde{k}}+\widetilde{G}_U(s)}+\cfrac{1}{\cfrac{1}{\widetilde{b}s}+\widetilde{G}_L(s)}}.
    \label{eq:treeRecursive}
\end{equation}
Moving one generation deeper, the transfer function from the input force to the relative distance between $x_{2,1}$ and $x_{\text{last}}$ is  $\widetilde{G}_U(s)$; similarly, $\widetilde{G}_L(s)$ is the transfer function between $x_{2,2}$ and $x_{\text{last}}$. The spring constant connecting $x_{1,1}$ to $x_{2,1}$ is denoted by $\widetilde{k}$, and $\widetilde{b}$ denotes the damper constant connecting $x_{1,1}$ to $x_{2,2}$. Fig.~\ref{fig:recursive} illustrates the meaning of above elements.

\subsection{Undamaged versus Damaged Tree}
\label{sec:undVsD}

Throughout this paper, we call the tree model \emph{undamaged} when all spring constants and all damper constants are same respectively, that is $k_{g,n}=k$ and $b_{g,n}=b$ for all $g=1,2,\dots$ and $n=1,2,\dots,2^{g-1}$.
Moreover, for each damage case, we assume that there is either only one spring or only one damper having a constant different from its corresponding undamaged value. We further assume that the damaged spring (damper) constant $k_d$ ($b_d$) is related to its corresponding undamaged value $k$ ($b$) by a factor of $\epsilon$, \textit{i.e.},
\begin{equation*}
    k_d=k\cdot\epsilon\text{  or  }b_d=b\cdot\epsilon,
\end{equation*}
where $\epsilon$ is called the damage amount. In addition, we also assume that the damaged component is weaker than the undamaged ones, \textit{i.e.}, $0<\epsilon<1$.

As shown in \cite{goodwine2014modeling}, for the undamaged case, since all the springs (dampers) have the same constant and the number of generations is infinite, all those three transfer functions $\widetilde{G}(s)$, $\widetilde{G}_U(s)$, and $\widetilde{G}_L(s)$ in Eq.~(\ref{eq:treeRecursive}) have the same expression, which is called the undamaged transfer function $G_\infty(s)$. Therefore, for the undamaged case, we can replace ($\widetilde{G}(s)$, $\widetilde{G}_U(s)$, $\widetilde{G}_L(s)$, $\widetilde{k}$, $\widetilde{b}$) with ($G_\infty(s)$, $G_\infty(s)$, $G_\infty(s)$, $k$, $b$) in Eq.~(\ref{eq:treeRecursive}). Then, that becomes an equation with only one unknown, namely $G_\infty(s)$, which yields that the transfer function from the input force $f(t)$ to the relative distance between $x_{1,1}(t)$ and $x_{\text{last}}(t)$ for the undamaged tree is given by
\begin{equation}
    G_\infty(s)=\frac{X_{1,1}(s)-X_{\text{last}}(s)}{F(s)}=\frac{1}{\sqrt{kbs}}.
    \label{eq:undTreeTF}
\end{equation}

\subsection{Recurrence Formula}
\label{sec:equation1}

Eq.~(\ref{eq:treeRecursive}) can be viewed as a mapping from ($\widetilde{G}_U(s)$, $\widetilde{G}_L(s)$) to $\widetilde{G}(s)$, which essentially builds up the entire tree generation by generation. This is always the case no matter whether the tree model is undamaged or damaged.

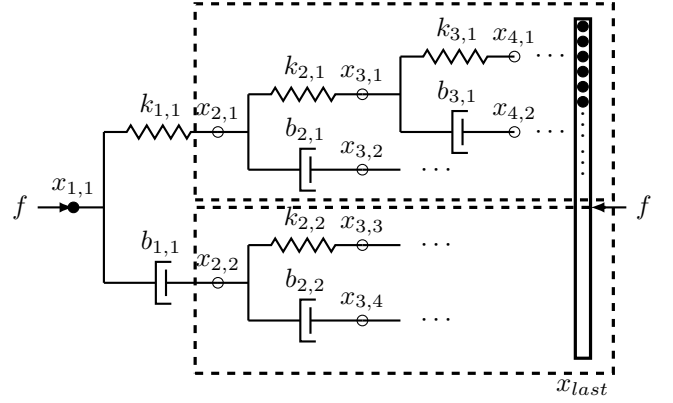
\begin{figure}
\centering
\begin{tikzpicture}
\tikzstyle{spring}=[thick,decorate,decoration={zigzag,pre length=0.3cm,post length=0.3cm,segment length=6}]
\tikzstyle{damper}=[thick,decoration={markings,mark connection node=dmp,mark=at position 0.5 with 
   {
    \node (dmp) [thick,inner sep=0pt,transform shape,rotate=-90,minimum
 width=15pt,minimum height=3pt,draw=none] {};
    \draw [thick] ($(dmp.north east)+(2pt,0)$) -- (dmp.south east) -- (dmp.south
 west) -- ($(dmp.north west)+(2pt,0)$);
    \draw [thick] ($(dmp.north)+(0,-5pt)$) -- ($(dmp.north)+(0,5pt)$);
   }
 },decorate]
 
\node at (0,0) (leftF) {$f$};
\filldraw[black] (0.7,0) circle (2pt) node [above] {$x_{1,1}$};
\draw[thick, -latex] (leftF.east) -- (0.7,0);
\draw[thick] (0.7,0) -- (1.1,0);
\draw[thick] (1.1,1) -- (1.1,-1);
\draw[spring] (1.1,1) -- (2.6,1) node [midway,above=1pt] {$k_{1,1}$};
\draw[damper] (1.1,-1) -- (2.6,-1) node [midway,above=6pt] {$b_{1,1}$};
\draw (2.6,1) circle (2pt) node [above] {$x_{2,1}$};
\draw (2.6,-1) circle (2pt) node [above] {$x_{2,2}$};
\draw[thick] (2.6,1) -- (3,1);
\draw[thick] (2.6,-1) -- (3,-1);
\draw[thick] (3,1.5) -- (3,0.5);
\draw[thick] (3,-0.5) -- (3,-1.5);
\draw[spring] (3,1.5) -- (4.5,1.5) node [midway,above=1pt] {$k_{2,1}$};
\draw[damper] (3,0.5) -- (4.5,0.5) node [midway,above=6pt] {$b_{2,1}$};
\draw[spring] (3,-0.5) -- (4.5,-0.5) node [midway,above=1pt] {$k_{2,2}$};
\draw[damper] (3,-1.5) -- (4.5,-1.5) node [midway,above=6pt] {$b_{2,2}$};
\draw (4.5,1.5) circle (2pt) node [above] {$x_{3,1}$};
\draw (4.5,0.5) circle (2pt) node [above] {$x_{3,2}$};
\draw (4.5,-0.5) circle (2pt) node [above] {$x_{3,3}$};
\draw (4.5,-1.5) circle (2pt) node [above] {$x_{3,4}$};
\draw[thick] (4.5,1.5) -- (5,1.5);
\draw[thick] (4.5,0.5) -- (5,0.5);
\draw[thick] (4.5,-0.5) -- (5,-0.5);
\draw[thick] (4.5,-1.5) -- (5,-1.5);
\draw[thick] (5,2) -- (5,1);
\draw[spring] (5,2) -- (6.5,2) node [midway,above=1pt] {$k_{3,1}$};
\draw[damper] (5,1) -- (6.5,1) node [midway,above=6pt] {$b_{3,1}$};
\draw (6.5,2) circle (2pt) node [above] {$x_{4,1}$};
\draw (6.5,1) circle (2pt) node [above] {$x_{4,2}$};
\node at (7,2) {$\cdots$};
\node at (7,1) {$\cdots$};
\node at (5.5,0.5) {$\cdots$};
\node at (5.5,-0.5) {$\cdots$};
\node at (5.5,-1.5) {$\cdots$};
\draw[black, very thick] (7.3,2.5) rectangle (7.5,-2) node [midway,below=2.4cm] {$x_{last}$};
\filldraw[black] (7.4,2.4) circle (2pt);
\filldraw[black] (7.4,2.2) circle (2pt);
\filldraw[black] (7.4,2) circle (2pt);
\filldraw[black] (7.4,1.8) circle (2pt);
\filldraw[black] (7.4,1.6) circle (2pt);
\filldraw[black] (7.4,1.4) circle (2pt);
\node at (7.4,1.2) {$\vdots$};
\node at (7.4,0.7) {$\vdots$};
\node at (8.2,0) (rightF) {$f$};
\draw[thick, -latex] (rightF.west) -- (7.5,0);
\draw[black, very thick, dashed] (2.3,2.7) rectangle (7.8,0.1);
\draw[black, very thick, dashed] (2.3,0) rectangle (7.8,-2.2);
\end{tikzpicture}
\caption{The tree model is self-similar. When undamaged, both two sub-networks encircled by boxes are same as the entire network.}
\label{fig:selfSim}
\end{figure}

The above sub-section presents the existing literature which shows that the undamaged tree's transfer function $G_\infty(s)$ can be obtained by replacing ($\widetilde{G}(s)$, $\widetilde{G}_U(s)$, $\widetilde{G}_L(s)$, $\widetilde{k}$, $\widetilde{b}$) with ($G_\infty(s)$, $G_\infty(s)$, $G_\infty(s)$, $k$, $b$) in Eq.~(\ref{eq:treeRecursive}). Essentially, this means that for the undamaged tree model, the transfer function between $x_{1,1}$ and $x_{\text{last}}$ is the same as the one between $x_{2,1}$ and $x_{\text{last}}$, as well as the one between $x_{2,2}$ and $x_{\text{last}}$. Fig.~\ref{fig:selfSim} illustrates the nature of this type of self-similarity.

In a similar manner, taking advantage of self-similarity, every damage case can also be computed by using Eq.~(\ref{eq:treeRecursive}) repeatedly. For example, if we replace ($\widetilde{G}_U(s)$, $\widetilde{G}_L(s)$, $\widetilde{k}$, $\widetilde{b}$) with ($G_\infty(s)$, $G_\infty(s)$, $k\epsilon$, $b$) in Eq.~(\ref{eq:treeRecursive}), the resultant $\widetilde{G}(s)$ is the damaged transfer function when $k_{1,1}$ is damaged, which we denote as $G_{k_{1,1}}(s)$. Then, if we further replace ($\widetilde{G}_U(s)$, $\widetilde{G}_L(s)$, $\widetilde{k}$, $\widetilde{b}$) with ($G_{k_{1,1}}(s)$, $G_\infty(s)$, $k$, $b$) in Eq.~(\ref{eq:treeRecursive}) again, the resultant $\widetilde{G}(s)$ is the damaged transfer function when the damage occurs at $k_{2,1}$.

Note that, after taking the above procedure repeatedly, the transfer function for the entire damaged tree is very complicated. In fact, both transfer functions $\widetilde{G}_U(s)$ and $\widetilde{G}_L(s)$ have the same formula as Eq.~(\ref{eq:treeRecursive}) due to the self-similarity. Therefore, when only integer-order calculus can be used, the transfer function for the entire tree inevitably becomes a complicated infinite continued fraction, which consists of infinitely many copies of Eq.~(\ref{eq:treeRecursive}):
\begin{equation}
    G(s)=\begin{bmatrix}\cfrac{1}{\cfrac{1}{k_{1,1}}+\cfrac{1}{\cfrac{1}{\cfrac{1}{k_{2,1}}+\ddots}+\cfrac{1}{\cfrac{1}{b_{2,1}s}+\ddots}}}\\
    +\cfrac{1}{\cfrac{1}{b_{1,1}s}+\cfrac{1}{\cfrac{1}{\cfrac{1}{k_{2,2}}+\ddots}+\cfrac{1}{\cfrac{1}{b_{2,2}s}+\ddots}}}\end{bmatrix}^{-1}.
    \label{eq:treeInt}
\end{equation}
Actually, the undamaged transfer function, Eq.~(\ref{eq:undTreeTF}), can be viewed as the limit to which Eq.~(\ref{eq:treeInt}) converges when $k_{1,1}=k_{2,1}=\cdots=k$ and $b_{1,1}=b_{2,1}=\cdots=b$. Obviously, Eq.~(\ref{eq:undTreeTF}) is much more concise, yet it involves fractional-order derivatives. Our research of modeling damaged trees presented in this paper extends that idea and proves the limit to which Eq.~(\ref{eq:treeInt}) converges, even when there exists one of those constants being different from its nominal value.

\subsection{Fractional-Order Calculus}

Recall that, $sX(s)$ in the frequency domain corresponds to $\frac{d}{dt}x(t)$ in the time domain, assuming zero initial conditions. Fractional-order calculus extends that idea. For example, from the undamaged transfer function, Eq.~(\ref{eq:undTreeTF}), we can conclude that
\begin{equation*}
    \sqrt{kbs}\left(X_{1,1}(s)-X_{\text{last}(s)}\right)=F(s),
\end{equation*}
which leads to its equation of motion
\begin{equation*}
    \sqrt{kb}\frac{d^{0.5}}{dt^{0.5}}\left(x_{1,1}(t)-x_{\text{last}}(t)\right)=f(t).
\end{equation*}

\section{Main Result}
\label{sec:mathAnalysis}

The main objective of this paper is to model a damaged tree in a way which is easier for analysis and control. Although Eq.~(\ref{eq:treeRecursive}) can construct the transfer function for all damaged cases, it is not particularly useful because it is not a concise expression as we see in the previous section. What we are looking for is to write down a damaged tree's analytical transfer function as concise as Eq.(\ref{eq:undTreeTF}).

The main result of this paper is that we actually can write down a damaged tree's transfer function as
\begin{equation*}
    G_l(s)=G_\infty(s)\Delta_l(s),
\end{equation*}
where the disturbance $\Delta_l(s)$ is well structured and can be determined completely by the damage amount $\epsilon$ of a damaged component $l$. As we are going to show later in Section \ref{sec:discussion}, those two features are the key points which make such way of modeling useful in different applications.

\subsection{$\Delta(s)$ Is Well-structured}
\label{sec:structured}

\textbf{Claim:} For each damage case as defined in Section \ref{sec:undVsD}, its damaged transfer function $G_l(s)$ from the input force to the relative distance between $x_{1,1}$ and $x_{\text{last}}$ can always be modeled as a fractional-order nominal plant with a fractional-order multiplicative disturbance, that is,
\begin{equation}
    G_l(s)=G_\infty(s)\Delta_l(s),
    \label{eq:damaged}
\end{equation}
where $G_\infty(s)$ is the undamaged transfer function defined by Eq.~(\ref{eq:undTreeTF}). Moreover, $\Delta(s)$ is structured as
\begin{equation}
    \Delta_l(s)=\frac{N(s)}{D(s)}=\frac{\prod_{j=1}^{2g}(s^{\frac{1}{2}}+z_j)}{\prod_{j=1}^{2g}(s^{\frac{1}{2}}+p_j)}
    \label{eq:delta}
\end{equation}
where $g$ denotes the $g$-th generation at which the damaged component $l$ locates, and $-z_j$ and $-p_j$ are called as half-order zeros and poles. In addition, $z_1$ is fixed at $\sqrt{\frac{k}{b}}$ no matter where the damage locates and no matter how large the damage amount $\epsilon$ is.

Note that Eq.~(\ref{eq:damaged}) is the exact model of a damaged tree's frequency response, rather than a model which we subjectively pick to approximate that frequency response. As we will see in the following proof, Eq.~(\ref{eq:damaged}) is the analytical result in the form of a rational expression from repeating using the recurrence formula~(\ref{eq:treeRecursive}) to obtain a damaged tree's frequency response according to the logic described in Section~\ref{sec:equation1}. In other words, Eq.~(\ref{eq:damaged}) is the analytical limit point to which the continued fraction~(\ref{eq:treeInt}) converges when one of those constants is different from its nominal value.

\textbf{Proof:} The proof is by induction. 

\underline{Base Case:} First, we prove that when the damage happens at the first generation, \textit{i.e.}, either $k_{1,1}=k_d=k\cdot\epsilon$ or $b_{1,1}=b_d=b\cdot\epsilon$, the damaged transfer function $G_l(s)$ satisfies the above claim.

For the reason explained in Section \ref{sec:equation1}, the damaged transfer function for the case when the damaged component is $l=k_{1,1}$ can be obtained by replacing ($\widetilde{G}_U(s)$, $\widetilde{G}_L(s)$, $\widetilde{k}$, $\widetilde{b}$) with ($G_\infty(s)$, $G_\infty(s)$, $k\cdot\epsilon$, $b$) in Eq.~(\ref{eq:treeRecursive}), which, when simplified gives
\begin{align}
    &G_{k_{1,1}}(s)=G_\infty(s)\nonumber\\
    &\cdot\frac{\left(s^\frac{1}{2}+\sqrt{\frac{k}{b}}\right)\left(s^\frac{1}{2}+\epsilon\sqrt{\frac{k}{b}}\right)}{\left(s^\frac{1}{2}+\epsilon\sqrt{\frac{k}{b}}+\sqrt{\frac{\epsilon(\epsilon-1)k}{b}}\right)\left(s^\frac{1}{2}+\epsilon\sqrt{\frac{k}{b}}-\sqrt{\frac{\epsilon(\epsilon-1)k}{b}}\right)}.\label{eq:k11}
\end{align}

Similarly, the damaged transfer function $G(s)$ for the damage at $b_{1,1}$ is obtained by replacing ($\widetilde{G}_U(s)$, $\widetilde{G}_L(s)$, $\widetilde{k}$, $\widetilde{b}$) with ($G_\infty(s)$, $G_\infty(s)$, $k$, $b\cdot\epsilon$) in Eq.~(\ref{eq:treeRecursive}), which is
\begin{align}
    &G_{b_{1,1}}(s)=G_\infty(s)\nonumber\\
    &\cdot\frac{\left(s^\frac{1}{2}+\sqrt{\frac{k}{b}}\right)\left(s^\frac{1}{2}+\frac{1}{\epsilon}\sqrt{\frac{k}{b}}\right)}{\left(s^\frac{1}{2}+\sqrt{\frac{k}{b}}+\sqrt{\frac{(\epsilon-1)k}{\epsilon b}}\right)\left(s^\frac{1}{2}+\sqrt{\frac{k}{b}}-\sqrt{\frac{(\epsilon-1)k}{\epsilon b}}\right)}.\label{eq:b11}
\end{align}

Therefore, the base case, where the damaged component is either $l=k_{1,1}$ or $l=b_{1,1}$, satisfies the above claim.

\underline{Inductive Step:} Assume that the damaged transfer function $\overline{G}(s)$ when the damage happens at $g$-th generation, \textit{i.e.}, either at $k_{g,n}$ or at $b_{g,n}$ ($g=1,2,3,\dots$ and $n=1,2,3,\dots,2^{g-1}$), satisfies the above claim. That is,
\begin{align*}
    \overline{G}(s)&=G_\infty(s)\overline{\Delta}(s)=G_\infty(s)\frac{N(s)}{D(s)}\\
    &=G_\infty(s)\frac{\prod_{j=1}^{2g}(s^\frac{1}{2}+z_i)}{\prod_{j=1}^{2g}(s^\frac{1}{2}+p_i)},
\end{align*}
where $z_1=\sqrt{\frac{k}{b}}$. Note that the base case satisfies this assumption and corresponds to the case where $g=1$.

When that damage goes one generation deeper, \textit{i.e.}, from the $g$-th generation to the $(g+1)$-th generation, there exist two different cases.

Type 1: When the damage is in the upper half of the $(g+1)$-th generation, \textit{i.e.}, when the damaged component is either $l=k_{g+1,n}$ or $l=b_{g+1,n}$ where $n=1,2,3,\dots,2^{g-1}$, the transfer function $G_l(s)$ can be obtained by replacing ($\widetilde{G}_U(s)$, $\widetilde{G}_L(s)$, $\widetilde{k}$, $\widetilde{b}$) with ($G(s)\frac{N(s)}{D(s)}$, $G_\infty(s)$, $k$, $b$) in Eq.~(\ref{eq:treeRecursive}).
After simplifying, we obtain
\begin{equation}
    G_l(s)=G_\infty(s)\frac{\left(s^\frac{1}{2}+\sqrt{\frac{k}{b}}\right)\left(s^\frac{1}{2}D(s)+\sqrt{\frac{k}{b}}N(s)\right)}{sD(s)+\sqrt{\frac{k}{b}}s^\frac{1}{2}\left(N(s)+D(s)\right)+\frac{k}{b}D(s)}.
    \label{eq:proofIS1}
\end{equation}

Type 2: The other case is when the damage is in the lower half of the $(g+1)$-th generation, \textit{i.e.}, when the damaged component is either $l=k_{g+1,n}$ or $l=b_{g+1,n}$ where $n=2^{g-1}+1,\dots,2^g$. The transfer function $G_l(s)$ can be obtained by replacing ($\widetilde{G}_U(s)$, $\widetilde{G}_L(s)$, $\widetilde{k}$, $\widetilde{b}$) with ($G_\infty(s)$, $G_\infty(s)\frac{N(s)}{D(s)}$, $k$, $b$) in Eq.~(\ref{eq:treeRecursive}).
Again, after simplifying, we can obtain
\begin{equation}
    G_l(s)=G_\infty(s)\frac{\left(s^\frac{1}{2}+\sqrt{\frac{k}{b}}\right)\left(s^\frac{1}{2}N(s)+\sqrt{\frac{k}{b}}D(s)\right)}{sD(s)+\sqrt{\frac{k}{b}}s^\frac{1}{2}\left(N(s)+D(s)\right)+\frac{k}{b}D(s)}.
    \label{eq:proofIS2}
\end{equation}

Note that, for both Eq.~(\ref{eq:proofIS1}) and Eq.~(\ref{eq:proofIS2}), $s^\frac{1}{2}$ can be factored out of both the numerator and denominator such that the new multiplicative disturbance $\Delta_l(s)$ is of the form
\begin{equation*}
    \Delta_l(s)=\frac{\prod_{j=1}^{2(g+1)}\left(s^\frac{1}{2}+z_j\right)}{\prod_{j=1}^{2(g+1)}\left(s^\frac{1}{2}+p_j\right)},
\end{equation*}
where $z_1$ is still fixed at $\sqrt{\frac{k}{b}}$. Although the closed-form expression for those half-order zeros and poles are complicated and difficult to compute by hand, their numerical values can be easily computed by a nonlinear equation solver. \textit{Q.E.D.}

\subsection{$\Delta_l(s)$ Depends on $\epsilon$ Only at Each $l$}
\label{sec:deltaEpsilon}

\begin{figure}
\centering
\includegraphics[width=0.48\textwidth]{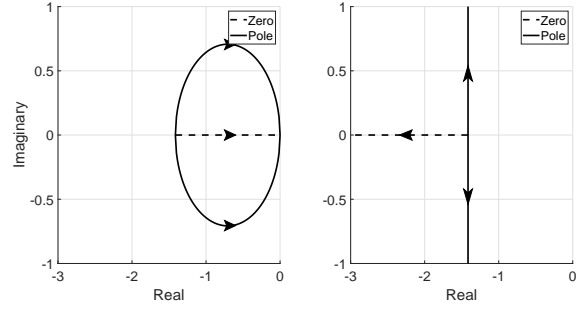}
\caption{Half-order zero-pole locus when the damage occurs at the first generation, and the damage amount $\epsilon$ varies from $1$ to $0$. When $\epsilon=1$, all half-order zeros and poles are at $-\sqrt{\frac{k}{b}}$. (For this plot, $k=2$ and $b=1$.) Left: $l=k_{1,1}$. Right: $l=b_{1,1}$.}
\label{fig:locus}
\end{figure}

When the damage happens at the first generation, the relation between $\Delta_l(s)$ and $\epsilon$ can be expressed in closed-form as shown in Eq.~(\ref{eq:k11}) and Eq.~(\ref{eq:b11}). Fig.~\ref{fig:locus} shows the locus for those half-order zeros and poles when the damage happens at the first generation, and when the damage amount $\epsilon$ varies from $1$ (no damage) to $0$ (complete damage).

\begin{figure}
\centering
\includegraphics[width=0.48\textwidth]{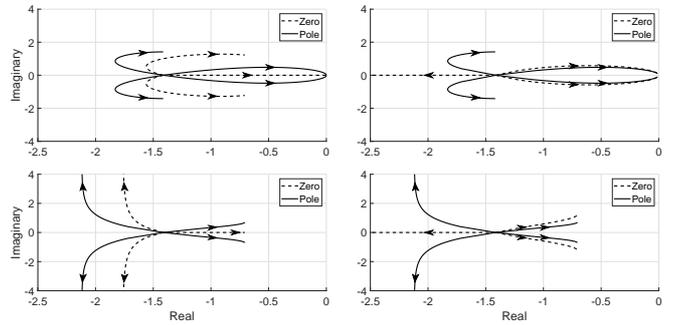}
\caption{Half-order zero-pole locus when the damage occurs at the second generation, and the damage amount $\epsilon$ varies from $1$ to $0$. When $\epsilon=1$, all half-order zeros and poles are at $-\sqrt{\frac{k}{b}}$. (For this plot, $k=2$ and $b=1$.) Upper left: $l=k_{2,1}$. Upper right: $l=k_{2,2}$. Lower left: $l=b_{2,1}$. Lower right: $l=b_{2,2}$.}
\label{fig:locus2}
\end{figure}

For all the other damage locations deeper into the network than the first generation, the relation between $\Delta_l(s)$ and $\epsilon$ cannot be easily expressed in a closed form. However, as was observed in Section \ref{sec:structured}, we can still obtain the locus of those half-order zeros and poles by using a nonlinear equation solver. Fig.~\ref{fig:locus2} shows the locus for those half-order zeros and poles, which are built up numerically, when the damage happens at the second generation, and when the damage amount $\epsilon$ varies from $1$ (no damage) to $0$ (complete damage).

Since it is possible to get this kind of locus for each damaged component, $\Delta_l(s)$ clearly has only one degree of freedom, namely $\epsilon$, at each damaged component $l$. That is, as long as either one pole or one zero (other than $-z_1$ which always stays at $-\sqrt{\frac{k}{b}}$) is known, all the other zeros and poles can be determined through $\epsilon$, thus $\Delta_l(s)$ is determined thereby.

\begin{figure}
\centering
\includegraphics[width=0.48\textwidth]{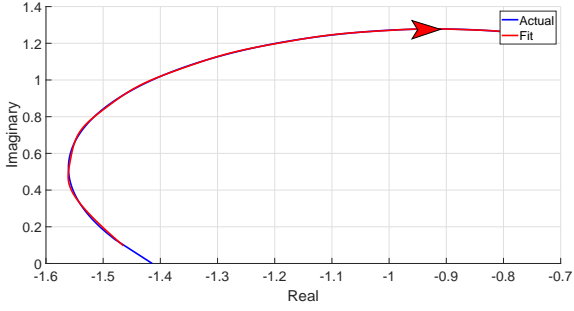}
\caption{The result for fitting one of the half-order zeros as a function of $\epsilon$ when the damage happens at $k_{2,1}$.}
\label{fig:locusFit}
\end{figure}

The exact relation between $\Delta_l(s)$ and $\epsilon$ is difficult to obtain unless the damage occurs at the first generation. However, we can compute that numerically by using some fitting techniques to the real and imaginary part of those half-order zero-pole locus. For example, Fig.~\ref{fig:locusFit} shows the least-square fitting result for one of the half-order zeros when the damage happens at $k_{2,1}$. That fitting uses the polynomial basis from $\epsilon^0$ to $\epsilon^{17}$. Note that Fig.~\ref{fig:locusFit} is merely an initial result showing that we are able to numerically obtain the relation between $\Delta_l(s)$ and $\epsilon$ when the damage occurs deeper than the first generation. In the future, we are going to consider using other regression techniques to obtain a better fitting result.

\subsection{Utility of These Results}
\begin{figure}
\centering
\includegraphics[width=0.48\textwidth]{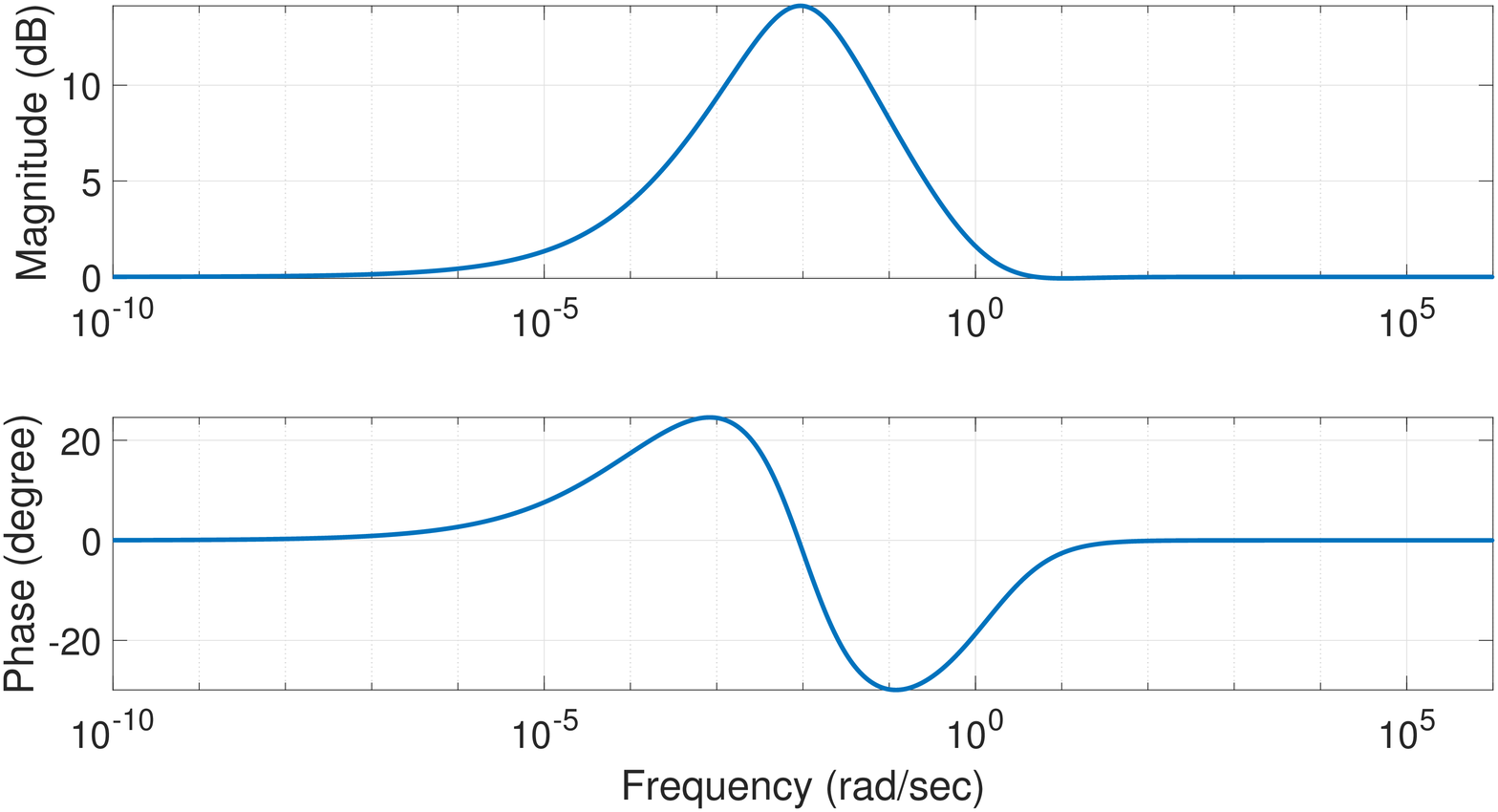}
\caption{Bode plot of $\Delta_{k_{2,1}}(s)$ when the damage amount $\epsilon=0.01$.}
\label{fig:deltaK21Bode}
\end{figure}

Now we describe why knowing that $\Delta_l(s)$ depends on $\epsilon$ solely is important and useful.
Fig.~\ref{fig:deltaK21Bode} shows the Bode plot of $\Delta_{k_{2,1}}(s)$ when the damage amount $\epsilon=0.01$.

Consider the problem of identifying $\Delta_{k_{2,1}}(s)$ given its Bode plot Fig.~\ref{fig:deltaK21Bode}. Assume that we know the structure of $\Delta_{k_{2,1}}(s)$ is defined by Eq.~(\ref{eq:delta}), but we do not know the fact that $\Delta_{k_{2,1}}(s)$ is completely determined by the damage amount $\epsilon$. Therefore, following the identification procedure in Section 4.1 of \cite{leyden2018system}, we can construct a nonlinear optimization problem
\begin{equation}
    \min_{z_j,p_j}\sum\frac{\|\widetilde{\Delta}_{k_{2,1}}(s)-\Delta_{k_{2,1}}(s)\|}{\|\Delta_{k_{2,1}}(s)\|},
    \label{eq:idError}
\end{equation}
where
\begin{equation*}
    \widetilde{\Delta}_{k_{2,1}}(s)=\frac{\prod_{j=1}^{4}(s^{\frac{1}{2}}+z_j)}{\prod_{j=1}^{4}(s^{\frac{1}{2}}+p_j)},
\end{equation*}
such that
\begin{itemize}
    \item $z_1=\sqrt{\frac{k}{b}}$;
    \item $(z_j,z_{j+1}),(p_j,p_{j+1})$ are complex conjugate for $j=1,3$;
    \item $\text{Re}(z_j)\text{ and Re}(p_j)\geq0\text{ for all }j=1,\dots,4$.
\end{itemize}

\begin{figure}
\centering
\includegraphics[width=0.48\textwidth]{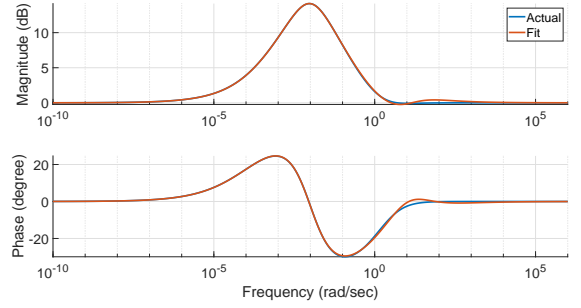}
\caption{Bode plot of the identified $\widetilde{\Delta}_{k_{2,1}}(s)$ when the damage amount $\epsilon=0.01$ if we don't use the knowledge that $\Delta_{k_{2,1}}(s)$ is a function of $\epsilon$.}
\label{fig:deltaK21Fit1}
\end{figure}

\begin{figure}
\centering
\includegraphics[width=0.48\textwidth]{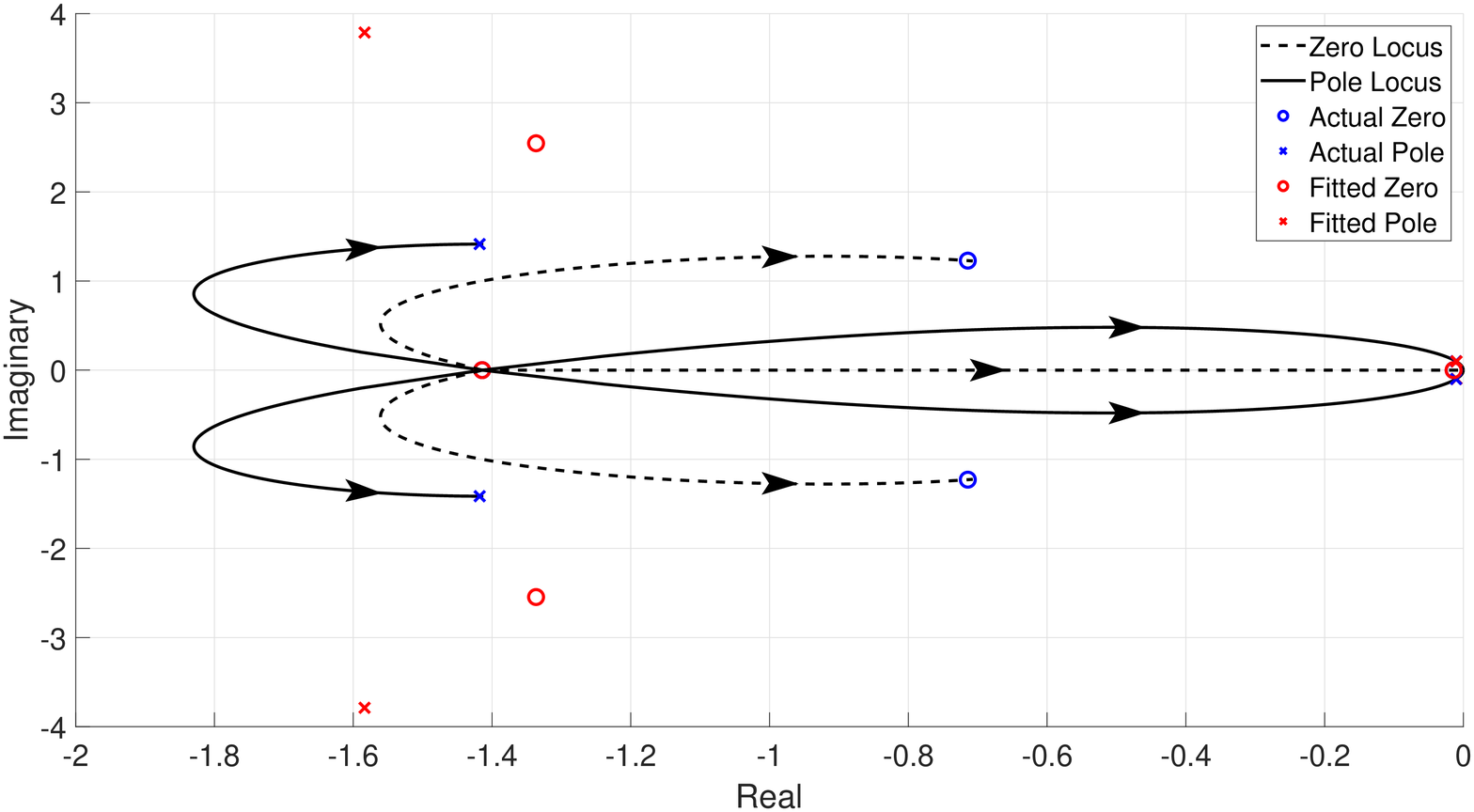}
\caption{Zero-pole locus plot for the identified $\widetilde{\Delta}_{k_{2,1}}(s)$ when the damage amount $\epsilon=0.01$ if we don't use the knowledge that $\Delta_{k_{2,1}}(s)$ is a function of $\epsilon$.}
\label{fig:deltaK21Fit1Locus}
\end{figure}

We solve the above nonlinear optimization problem using \texttt{fmincon()} in \texttt{MATLAB}, and draw the Bode plot of the fitted $\widetilde{\Delta}_{k_{2,1}}(s)$ in Fig.~\ref{fig:deltaK21Fit1}, which seems to be a promising result. However, the corresponding zeros and poles are far away from their locus, as shown in Fig.~\ref{fig:deltaK21Fit1Locus}.

In contrast, if we exploit the knowledge that $\Delta_{k_{2,1}}(s)$ is a function of $\epsilon$, we can rewrite the above nonlinear optimization problem as
\begin{equation*}
    \min_{\epsilon}\sum\frac{\|\widetilde{\Delta}_{k_{2,1}}(s)-\Delta_{k_{2,1}}(s)\|}{\|\Delta_{k_{2,1}}(s)\|}
\end{equation*}
such that
\begin{equation*}
    \widetilde{\Delta}_{k_{2,1}}(s)=\frac{\prod_{j=1}^{4}(s^{\frac{1}{2}}+z_j)}{\prod_{j=1}^{4}(s^{\frac{1}{2}}+p_j)}
\end{equation*}
and $z_j=z_j(\epsilon)$, $p_j=p_j(\epsilon)$ for all $j=1,\dots,4$. The functions $z_j(\epsilon)$ and $p_j(\epsilon)$ are known prior to solving this optimization problem by fitting the zero-pole locus as discussed before-mentioned.

\begin{figure}
\centering
\includegraphics[width=0.48\textwidth]{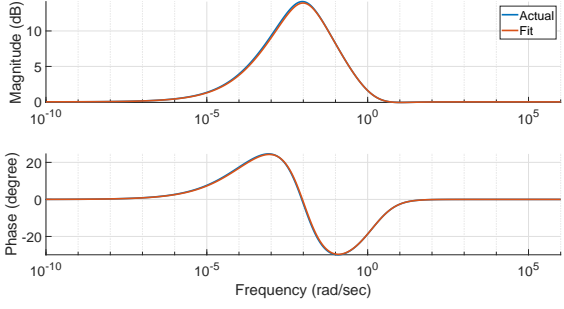}
\caption{Bode plot of the identified $\widetilde{\Delta}_{k_{2,1}}(s)$ when the damage amount $\epsilon=0.01$ if we use the knowledge that $\Delta_{k_{2,1}}(s)$ is a function of $\epsilon$.}
\label{fig:deltaK21Fit2}
\end{figure}

\begin{figure}
\centering
\includegraphics[width=0.48\textwidth]{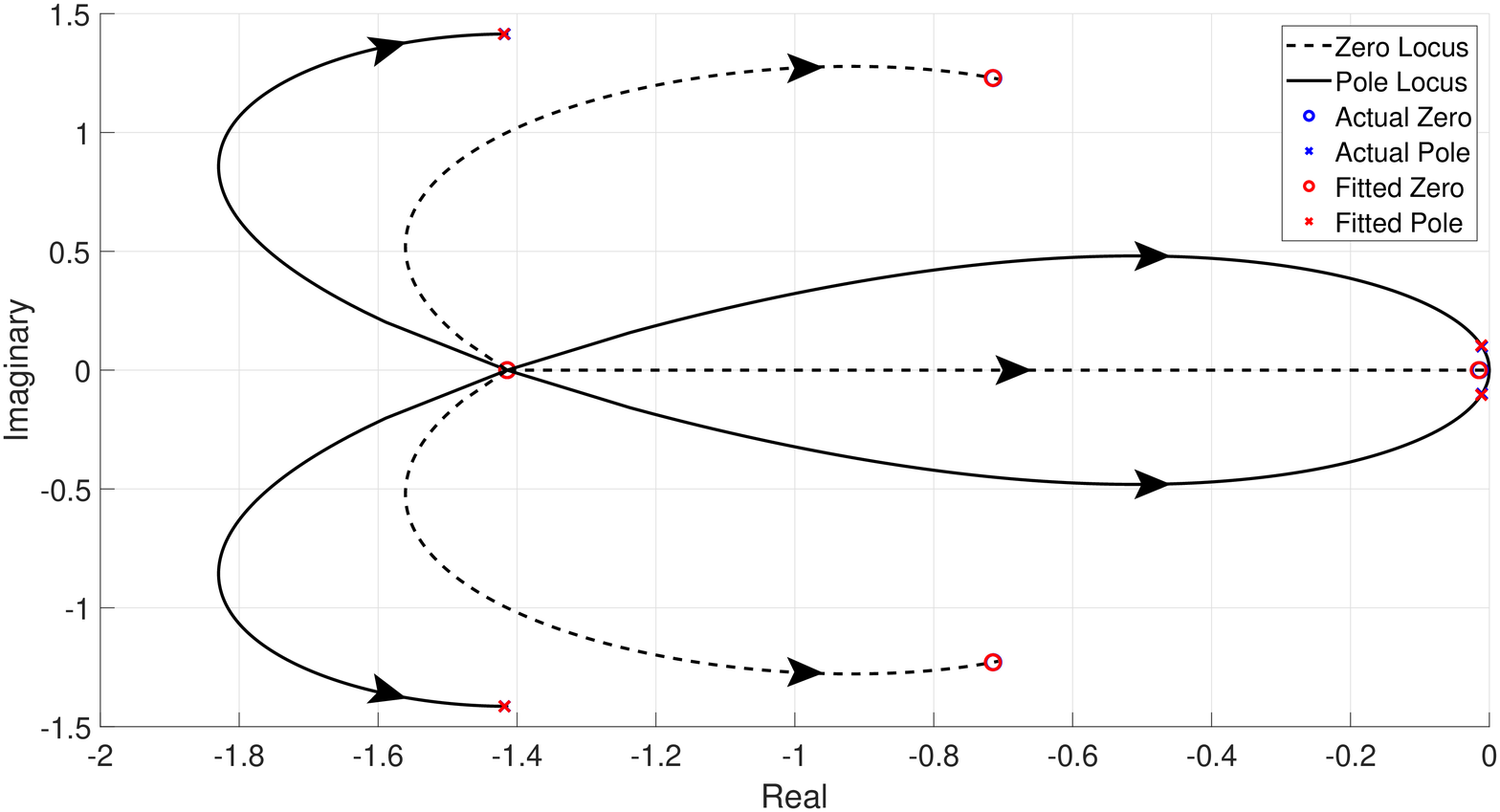}
\caption{Zero-pole locus plot for the identified $\widetilde{\Delta}_{k_{2,1}}(s)$ when the damage amount $\epsilon=0.01$ if we use the knowledge that $\Delta_{k_{2,1}}(s)$ is a function of $\epsilon$.}
\label{fig:deltaK21Fit2Locus}
\end{figure}

This time, \texttt{fmincon()} successfully identifies $\epsilon=0.01068$ from solving the above optimization problem. The resulting Bode plot and the zero-pole locus of the identified $\widetilde{\Delta}_{k_{2,1}}(s)$ are shown in Fig.~\ref{fig:deltaK21Fit2} and Fig.~\ref{fig:deltaK21Fit2Locus}, which are better than those results when the relation between $\Delta_{k_{2,1}}(s)$ and $\epsilon$ is not considered.

Note that the damage in the above example occurs at the second generation, which is a relatively easy case for identification. Therefore, it explains how necessary it is to consider binding all zeros and poles together through $\epsilon$. The main reason why doing so gives better identification result is that the respective contribution to the entire system of each zero and pole is obscure, and the frequency response of the whole system may be insensitive to some zeros and poles. By characterizing $\Delta_l(s)$ completely through $\epsilon$, we constrain those zeros and poles to move along their corresponding locus only and thus make the solution to the identification problem more accurate.

\section{DISCUSSION}
\label{sec:discussion}

In this section, we discuss three applications which will be benefited from the advantage of modeling a damaged network system as a product of a nominal plant and a fractional-order disturbance.

\subsection{Insights about How Different Types of Damage Affect The Network}
\label{sec:insights}

Modeling a damaged network in such way brings us insights about how that damage affects the network through a concise and fully characterized transfer function. For example, two observations for the damaged tree model are listed below. At present, we do not have clear explanations with regard to the physical meaning for those observations, but clearly those observations cannot be made without modeling the damaged tree model in such way.

\begin{enumerate}
\item Two damages at the corresponding location between the upper half and the lower half of a generation have the same poles. Therefore, how differently those two damages affect the network are completed determined by their respective zeros.
Specifically, the damage at $k_{g,n}$ ($b_{g,n}$) has the same poles as the damage at $k_{g,n+2^{g-2}}$ ($b_{g,n+2^{g-2}}$) for $n=1,\dots,2^{g-2}$ at the $g$-th generation. For example, as shown in Fig.~\ref{fig:locus2}, the damage at $k_{2,1}$ ($b_{2,1}$) has the same poles as the damage at $k_{2,2}$ ($b_{2,2}$). This observation can be confirmed by the fact that the denominators in Eq.~(\ref{eq:proofIS1}) and Eq.~(\ref{eq:proofIS2}) are same.

\item When the damage occurs at a spring, all poles for $\Delta_l(s)$ stay finite. In contrast, when a damage is to a damper, two poles for $\Delta_l(s)$ asymptotically approach infinity as the damage amount $\epsilon$ approaches $0$. This can be observed from both Fig.~\ref{fig:locus} and Fig.~\ref{fig:locus2}.
\end{enumerate}

\subsection{Robust Control}

The multiplicative disturbance, $G_\infty(s)\Delta_l(s)$, is a classical model in the robust control area. Therefore, modeling a damaged network in such way enables us to design a controller for that network by using some robust control methods.
For example, we may wish to design a robust controller for the damaged tree model such that the controller's performance is guaranteed under some uncertain damages whose location is unknown but the damage amount is bounded by $\epsilon_{\max}$.

\begin{figure}
\centering
\includegraphics[width=0.48\textwidth]{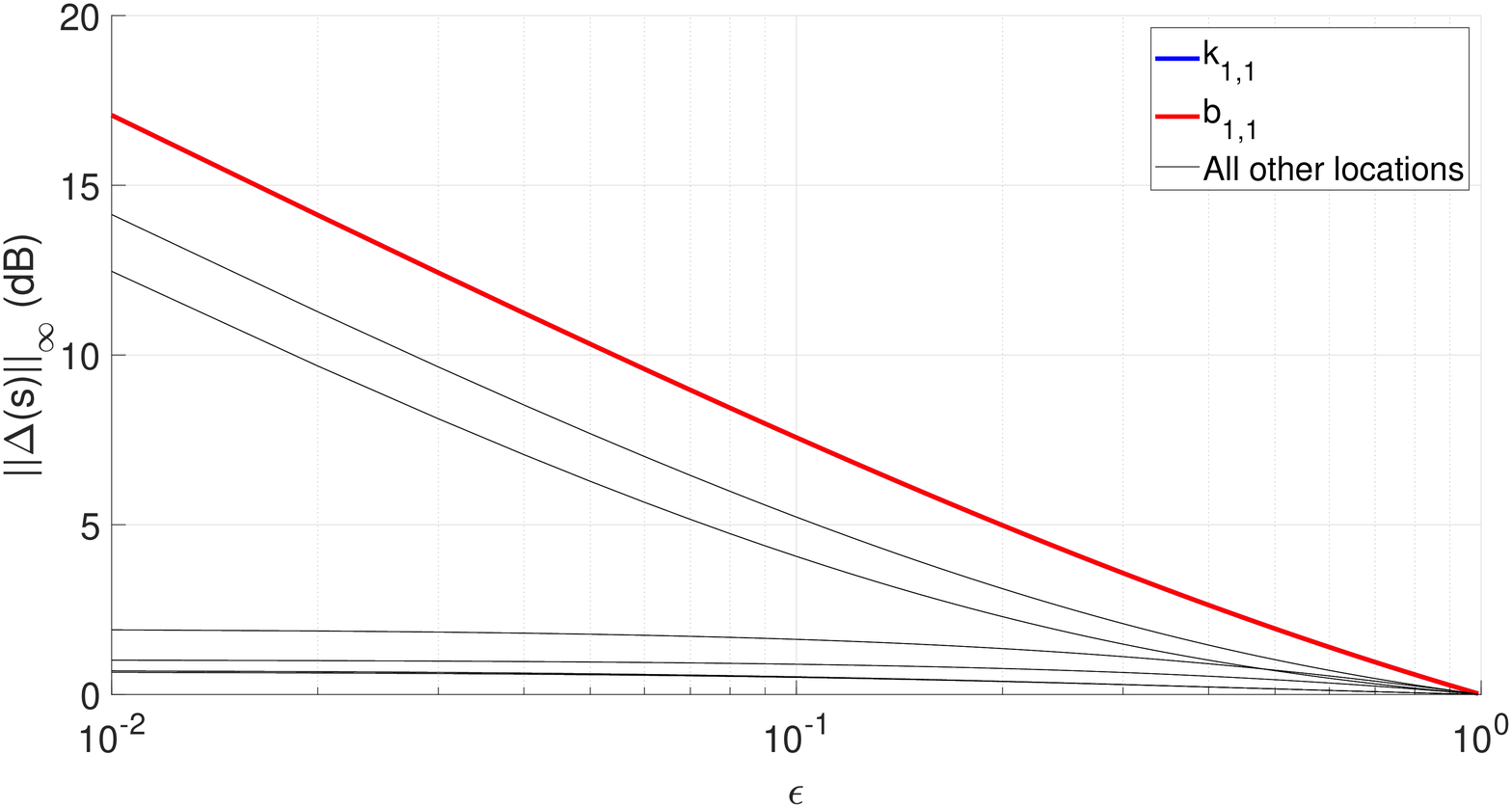}
\caption{$\|\Delta_l(s)\|_\infty$ at different damaged components \textit{versus} the damage amount $\epsilon$. Under the same damage amount $\epsilon$, $\|\Delta_l(s)\|_\infty$ at $k_{1,1}$ or $b_{1,1}$ sets the upper bound for all the other damaged components.}
\label{fig:deltaNorm}
\end{figure}

Fig.~\ref{fig:deltaNorm} plots $\|\Delta_l(s)\|_\infty$ \textit{versus} $\epsilon$, and it shows that, under the same damage amount, $\|\Delta_l(s)\|_\infty$ at $k_{1,1}$ or $b_{1,1}$ sets the upper bound for all the other damaged components. Conceptually, this observation is easy to understand, as the effect of a damage should be less and less noticeable when that damage goes deeper and deeper inside the network.

As a result, although we may not know exactly where a damage occurs, neither do we know the exact damage amount $\epsilon$, we can still bound $\|\Delta_l(s)\|_\infty$ analytically by replacing $\epsilon$ with $\epsilon_{\max}$ in Eq.~(\ref{eq:k11}) or Eq.~(\ref{eq:b11}) for all damage cases. 

\subsection{Identification for A Damaged Network}

As shown in Section \ref{sec:deltaEpsilon}, the identification for a damaged network becomes easier when we model it as $G_\infty(s)\Delta_l(s)$ where $\Delta_l(s)$ is well structured and can be completely characterized by the damage amount $\epsilon$ at each damage location. By taking advantage of the relation between $\Delta_l(s)$ and $\epsilon$, the identification problem becomes a single-variable optimization problem, which is easier to solve and, equally importantly, easier to visualize.

\begin{figure}
\centering
\includegraphics[width=0.48\textwidth]{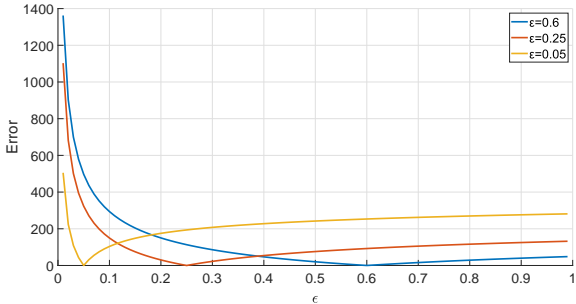}
\caption{The identification error, Eq.~(\ref{eq:idError}), \textit{versus} the damage amount $\epsilon$ when the damage occurs at $k_{1,1}$ and the actual damage $\Delta_l(s)$ is constructed by the damage amounts $\varepsilon=0.6$, $\varepsilon=0.25$ and $\varepsilon=0.05$.}
\label{fig:errorVsE}
\end{figure}

Fig.~\ref{fig:errorVsE} shows the objective function, namely the identification error $\sum\|\widetilde{\Delta_l}(s)-\Delta_l(s)\|/\|\Delta_l(s)\|$, \textit{versus} the damage amount $\epsilon$ when the damaged component is $l=k_{1,1}$ and the actual damage $\Delta_l(s)$ is constructed by the damage amounts $\varepsilon=0.6$, $\varepsilon=0.25$, and $\varepsilon=0.05$. From that figure, we see that the identification error reaches the minimum when the identified damage amount $\epsilon$ equals to the actual damage amount $\varepsilon$.

From these analyses, we can conclude the following.  First, we can confirm the fact that a severe damage is easier to identify, because the objective function, Eq.~(\ref{eq:idError}), is more sensitive when $\epsilon$ is close to $0$ compared to light damage (when $\epsilon$ is close to $1$). Therefore, we may be able to use such kind of plots to come up with an objective function which is more reasonable in the sense that it is more sensitive to light damages and thus makes light damages easier to identify. In addition, during the identification procedure, we can draw such type of plots at each optimization iteration so that if anything goes wrong during the identification, debugging might become easier.

\section{CONCLUSIONS AND FUTURE WORK}
\label{sec:conclusion}

In this paper, we show that a damage inside a large network can be modelled by a fractional-order nominal plant with a fractional-order multiplicative disturbance, $G_\infty(s)\Delta_l(s)$, where $\Delta_l(s)$ is well structured and can be completely characterized by the damage amount at each damaged component $l$. In addition, we also discuss three applications which take advantage of such way of modeling. First, this model brings us insights about how damage affects a network's behavior. Second, we can use such type of models to design a robust controller for a network under uncertain damages. Third, knowledge of modeling can also be used to identify an unknown damage given a network's frequency response. A damage identification algorithm is the focus of current research efforts and initial results indicate that the approach will be very effective.

In addition to damage identification, we are working to further the results in this paper in two ways. First, this paper limits the number of damaged components to one. Second, this paper only focuses on the tree model. Therefore, we are now considering approaches to model the tree with arbitrarily many damaged components. Besides, we are also working on developing a damage modeling algorithm for a class of networks.

%\begin{ack}
%Place acknowledgments here.
%\end{ack}

\bibliography{ifacconf}             % bib file to produce the bibliography
                                                     % with bibtex (preferred)
                                                   
%\begin{thebibliography}{xx}  % you can also add the bibliography by hand

%\bibitem[Able(1956)]{Abl:56}
%B.C. Able.
%\newblock Nucleic acid content of microscope.
%\newblock \emph{Nature}, 135:\penalty0 7--9, 1956.

%\bibitem[Able et~al.(1954)Able, Tagg, and Rush]{AbTaRu:54}
%B.C. Able, R.A. Tagg, and M.~Rush.
%\newblock Enzyme-catalyzed cellular transanimations.
%\newblock In A.F. Round, editor, \emph{Advances in Enzymology}, volume~2, pages
%  125--247. Academic Press, New York, 3rd edition, 1954.

%\bibitem[Keohane(1958)]{Keo:58}
%\newblock \emph{Power and Interdependence: World Politics in Transitions}.
%\newblock Little, Brown \& Co., Boston, 1958.

%\bibitem[Powers(1985)]{Pow:85}
%\newblock Is there a way out?
%\newblock \emph{Harpers}, pages 35--47, June 1985.

%\bibitem[Soukhanov(1992)]{Heritage:92}
%\newblock \emph{{The American Heritage. Dictionary of the American Language}}.
%\newblock Houghton Mifflin Company, 1992.

%\end{thebibliography}

%\appendix
%\section{A summary of Latin grammar}    % Each appendix must have a short title.
%\section{Some Latin vocabulary}              % Sections and subsections are supported  
                                                                         % in the appendices.
\end{document}